\documentclass[aps,prc,superscriptaddress,showpacs,floatfix,twocolumn]{revtex4}

\usepackage{graphicx}   

\begin{document}

\title{Centrality Dependence of Direct Photon Production in
$\sqrt{s_{_{NN}}}$~=~200~GeV Au+Au Collisions}

\newcommand{\abilene}{Abilene Christian University, Abilene, TX 79699, USA}
\newcommand{\acadsin}{Institute of Physics, Academia Sinica, Taipei 11529, Taiwan}
\newcommand{\banaras}{Department of Physics, Banaras Hindu University, Varanasi 221005, India}
\newcommand{\barc}{Bhabha Atomic Research Centre, Bombay 400 085, India}
\newcommand{\bnl}{Brookhaven National Laboratory, Upton, NY 11973-5000, USA}
\newcommand{\caucr}{University of California - Riverside, Riverside, CA 92521, USA}
\newcommand{\ciae}{China Institute of Atomic Energy (CIAE), Beijing, People's Republic of China}
\newcommand{\cns}{Center for Nuclear Study, Graduate School of Science, University of Tokyo, 7-3-1 Hongo, Bunkyo, Tokyo 113-0033, Japan}
\newcommand{\columbia}{Columbia University, New York, NY 10027 and Nevis Laboratories, Irvington, NY 10533, USA}
\newcommand{\dapnia}{Dapnia, CEA Saclay, F-91191, Gif-sur-Yvette, France}
\newcommand{\debrecen}{Debrecen University, H-4010 Debrecen, Egyetem t{\'e}r 1, Hungary}
\newcommand{\fsu}{Florida State University, Tallahassee, FL 32306, USA}
\newcommand{\gsu}{Georgia State University, Atlanta, GA 30303, USA}
\newcommand{\hiroshima}{Hiroshima University, Kagamiyama, Higashi-Hiroshima 739-8526, Japan}
\newcommand{\ihepprot}{Institute for High Energy Physics (IHEP), Protvino, Russia}
\newcommand{\isu}{Iowa State University, Ames, IA 50011, USA}
\newcommand{\jinrdubna}{Joint Institute for Nuclear Research, 141980 Dubna, Moscow Region, Russia}
\newcommand{\kaeri}{KAERI, Cyclotron Application Laboratory, Seoul, South Korea}
\newcommand{\kangnung}{Kangnung National University, Kangnung 210-702, South Korea}
\newcommand{\kek}{KEK, High Energy Accelerator Research Organization, Tsukuba-shi, Ibaraki-ken 305-0801, Japan}
\newcommand{\kfki}{KFKI Research Institute for Particle and Nuclear Physics (RMKI), H-1525 Budapest 114, POBox 49, Hungary}
\newcommand{\korea}{Korea University, Seoul, 136-701, Korea}
\newcommand{\kurchatov}{Russian Research Center ``Kurchatov Institute", Moscow, Russia}
\newcommand{\kyoto}{Kyoto University, Kyoto 606-8502, Japan}
\newcommand{\labllr}{Laboratoire Leprince-Ringuet, Ecole Polytechnique, CNRS-IN2P3, Route de Saclay, F-91128, Palaiseau, France}
\newcommand{\lawllnl}{Lawrence Livermore National Laboratory, Livermore, CA 94550, USA}
\newcommand{\losalamos}{Los Alamos National Laboratory, Los Alamos, NM 87545, USA}
\newcommand{\lpc}{LPC, Universit{\'e} Blaise Pascal, CNRS-IN2P3, Clermont-Fd, 63177 Aubiere Cedex, France}
\newcommand{\lund}{Department of Physics, Lund University, Box 118, SE-221 00 Lund, Sweden}
\newcommand{\muenster}{Institut f\"ur Kernphysik, University of Muenster, D-48149 Muenster, Germany}
\newcommand{\myongji}{Myongji University, Yongin, Kyonggido 449-728, Korea}
\newcommand{\nagasaki}{Nagasaki Institute of Applied Science, Nagasaki-shi, Nagasaki 851-0193, Japan}
\newcommand{\newmex}{University of New Mexico, Albuquerque, NM 87131, USA}
\newcommand{\nmsu}{New Mexico State University, Las Cruces, NM 88003, USA}
\newcommand{\ornl}{Oak Ridge National Laboratory, Oak Ridge, TN 37831, USA}
\newcommand{\orsay}{IPN-Orsay, Universite Paris Sud, CNRS-IN2P3, BP1, F-91406, Orsay, France}
\newcommand{\pnpi}{PNPI, Petersburg Nuclear Physics Institute, Gatchina, Russia}
\newcommand{\riken}{RIKEN (The Institute of Physical and Chemical Research), Wako, Saitama 351-0198, JAPAN}
\newcommand{\rikjrbrc}{RIKEN BNL Research Center, Brookhaven National Laboratory, Upton, NY 11973-5000, USA}
\newcommand{\saispbstu}{St. Petersburg State Technical University, St. Petersburg, Russia}
\newcommand{\saopaulo}{Universidade de S{\~a}o Paulo, Instituto de F\'{\i}sica, Caixa Postal 66318, S{\~a}o Paulo CEP05315-970, Brazil}
\newcommand{\seoulnat}{System Electronics Laboratory, Seoul National University, Seoul, South Korea}
\newcommand{\stonybrkc}{Chemistry Department, Stony Brook University, SUNY, Stony Brook, NY 11794-3400, USA}
\newcommand{\stonycrkp}{Department of Physics and Astronomy, Stony Brook University, SUNY, Stony Brook, NY 11794, USA}
\newcommand{\subatech}{SUBATECH (Ecole des Mines de Nantes, CNRS-IN2P3, Universit{\'e} de Nantes) BP 20722 - 44307, Nantes, France}
\newcommand{\tenn}{University of Tennessee, Knoxville, TN 37996, USA}
\newcommand{\titech}{Department of Physics, Tokyo Institute of Technology, Tokyo, 152-8551, Japan}
\newcommand{\tsukuba}{Institute of Physics, University of Tsukuba, Tsukuba, Ibaraki 305, Japan}
\newcommand{\vandy}{Vanderbilt University, Nashville, TN 37235, USA}
\newcommand{\waseda}{Waseda University, Advanced Research Institute for Science and Engineering, 17 Kikui-cho, Shinjuku-ku, Tokyo 162-0044, Japan}
\newcommand{\weizmann}{Weizmann Institute, Rehovot 76100, Israel}
\newcommand{\yonsei}{Yonsei University, IPAP, Seoul 120-749, Korea}
\affiliation{\abilene}
\affiliation{\acadsin}
\affiliation{\banaras}
\affiliation{\barc}
\affiliation{\bnl}
\affiliation{\caucr}
\affiliation{\ciae}
\affiliation{\cns}
\affiliation{\columbia}
\affiliation{\dapnia}
\affiliation{\debrecen}
\affiliation{\fsu}
\affiliation{\gsu}
\affiliation{\hiroshima}
\affiliation{\ihepprot}
\affiliation{\isu}
\affiliation{\jinrdubna}
\affiliation{\kaeri}
\affiliation{\kangnung}
\affiliation{\kek}
\affiliation{\kfki}
\affiliation{\korea}
\affiliation{\kurchatov}
\affiliation{\kyoto}
\affiliation{\labllr}
\affiliation{\lawllnl}
\affiliation{\losalamos}
\affiliation{\lpc}
\affiliation{\lund}
\affiliation{\muenster}
\affiliation{\myongji}
\affiliation{\nagasaki}
\affiliation{\newmex}
\affiliation{\nmsu}
\affiliation{\ornl}
\affiliation{\orsay}
\affiliation{\pnpi}
\affiliation{\riken}
\affiliation{\rikjrbrc}
\affiliation{\saispbstu}
\affiliation{\saopaulo}
\affiliation{\seoulnat}
\affiliation{\stonybrkc}
\affiliation{\stonycrkp}
\affiliation{\subatech}
\affiliation{\tenn}
\affiliation{\titech}
\affiliation{\tsukuba}
\affiliation{\vandy}
\affiliation{\waseda}
\affiliation{\weizmann}
\affiliation{\yonsei}
\author{S.S.~Adler}	\affiliation{\bnl}
\author{S.~Afanasiev}	\affiliation{\jinrdubna}
\author{C.~Aidala}	\affiliation{\bnl}
\author{N.N.~Ajitanand}	\affiliation{\stonybrkc}
\author{Y.~Akiba}	\affiliation{\kek} \affiliation{\riken}
\author{J.~Alexander}	\affiliation{\stonybrkc}
\author{R.~Amirikas}	\affiliation{\fsu}
\author{L.~Aphecetche}	\affiliation{\subatech}
\author{S.H.~Aronson}	\affiliation{\bnl}
\author{R.~Averbeck}	\affiliation{\stonycrkp}
% ...
\collaboration{PHENIX Collaboration} \noaffiliation

\date{\today}

\begin{abstract}

The first measurement of direct photons in Au+Au collisions at
$\sqrt{s_{_{NN}}} = 200$\,GeV is presented. The direct photon
signal is extracted as a function of the Au+Au collision centrality
and compared to NLO pQCD calculations.  The direct photon
yield is shown to scale with the number of nucleon-nucleon collisions
for all centralities.

\end{abstract}


\maketitle

One of the most exciting observations from experiments at the
Relativistic Heavy Ion Collider (RHIC) is the strong suppression
of the yield of hadrons at large transverse
momenta ($p_T > 2$ GeV/$c$) in central Au+Au collisions, as compared
to measured yields in $p+p$ collisions scaled by the number of
binary nucleon-nucleon collisions~\cite{ppg003,Adler:2002xw,ppg014,ppg026}.
Such quenching 
was predicted to result from the energy loss of hard-scattered
partons propagating through the high density
matter created in heavy ion collisions~\cite{Gyulassy:1990ye}.
It was later proposed that the observed hadron suppression
could be an initial-state effect due to saturation of the initial parton
distributions in large nuclei~\cite{Kharzeev:2002pc}. 
The high-$p_T$ hadron suppression was not observed 
in $d+$Au collisions~\cite{ppg028,Adams:2003im}.
This indicates that the suppression in Au+Au
collisions is due to the extended dense matter in the final state,
that is absent in $d+$Au collisions.

Measurement of direct photon production allows
more definitive discrimination between initial- and final-state
suppression due to the
fact that photons, once produced, are essentially unaffected by the
surrounding matter. Hence photons produced directly in initial parton
scatterings will not be quenched unless the initial parton distributions
are suppressed in the nucleus.
In fact, there may be additional direct photon yield 
in $AA$ collisions~\cite{Peitzmann:2001mz} 
due to various processes such as momentum broadening 
of the incoming partons, additional fragmentation 
contributions~\cite{Fries:2002kt,Zakharov:2004bi}, or
additional scatterings 
in the thermalizing dense matter of the final state.

This letter reports on direct photon production in Au+Au collisions at
$\sqrt{s_{_{NN}}} = 200$\,GeV with data taken by the PHENIX 
experiment~\cite{nim_phenix}
during the second RHIC run (2001). 
This analysis used the Beam-Beam Counters 
(BBC, $3.0<|\eta|<3.9$) and the Zero Degree Calorimeter (ZDC)
for trigger and event characterization, the Electromagnetic
Calorimeter (EMCal) in the two central arms
($|\eta|\leq0.35$) to measure the inclusive $\gamma$, 
$\pi^0$, and $\eta$ yields, and the tracking system of the
central arms to estimate the charged particle
contamination. The EMCal consists of two subsystems: six sectors of
lead-scintillator sandwich calorimeter (PbSc) and two sectors of lead-glass
Cherenkov calorimeter (PbGl).  Located at a radial distance of
about 5\,m each sector covers an azimuthal interval of $\Delta\phi \!  \approx
\! 22.5^{\circ}$. The fine segmentation of the EMCal ($\Delta \phi \! \times \!
\Delta \eta \sim 0.01 \! \times \!  0.01$) ensures that the two photons from a
decayed $\pi^{0}$ are well-resolved up to transverse momenta of ~15-20~GeV/$c$.

The event centrality was selected by cuts on the correlated distribution
of charged particles detected in the BBCs versus
energy measured in the ZDC detectors. A Glauber
model Monte Carlo combined with a simulation of the BBC and ZDC
responses gave an estimate of the associated number
of binary collisions ($N_{coll}$) and participating nucleons
($N_{part}$) for each centrality bin (values tabulated in
Ref.~\cite{ppg014}).

For this analysis a mimimum bias trigger sample of $30 \times
10^6$ events, also used for the previously published $\pi^0$
analysis~\cite{ppg014}, was combined with a Level-2 trigger event
sample equivalent to an additional $55 \times 10^6$ minimum bias
events. The Level-2 trigger sample was obtained by use of an EMCal
software trigger on highly energetic showers 
equivalent to the Level-1 hardware trigger  used
in Ref.~\cite{ppg024}.   The threshold energy of the
trigger was set at 3.5 GeV   
with a resulting trigger efficiency plateau at 100\% for single
photons above $p_T \approx 5\,\mbox{GeV}/c$ ($6.5\,\mbox{GeV}/c$)
for the PbSc (PbGl). The
normalization of the Level-2 data sample relative to the minimum bias
data sample is accurate to $2\%$. In the following, the minimum bias 
result refers to the combined Level-2 and minimum bias trigger samples without
selection on centrality.

The direct photon yield is extracted on a statistical basis, without isolation cuts,
by a comparison of the inclusive photon spectra to the expected
background from hadronic decays~\cite{Aggarwal:2000th,ppg049}
(mainly $\pi^0 \rightarrow 2\gamma$).
Photon-like clusters are identified in the EMCal by applying
appropriate Particle Identification (PID) cuts based on
time-of-flight and the shower profile.
The consistency of the final results obtained independently with the
PbSc and PbGl, and with different PID
cuts, including no PID cut, is used to check
the systematic error estimates.
The $\pi^0$ and $\eta$ yields are determined as described in
\cite{ppg014,ppgtbp} by an invariant mass analysis of photon pairs, with the
combinatorial background established by combining
uncorrelated photon pairs from different events.

The raw inclusive photon-candidate spectra must
be corrected for charged and neutral hadron contaminations not
removed by the PID cuts, as well as for photon conversions.
Charged contaminants
are identified by associating photon candidates in the EMCal with
charged hits in the pad chamber
(PC3) positioned directly in front of the EMCal. The charged contaminant
spectra are subtracted from the photon-candidate spectra.
The charged hadron contamination depends strongly on the PID
cut and increases significantly for $p_T < 3$~GeV/$c$ with a
contribution of ~4\% above 3~GeV/$c$ for the tightest PID cut. 
The contamination of neutral hadrons
(mainly anti-neutrons) is determined with a full GEANT simulation of
the detector response to neutrons and anti-neutrons with input spectra
based on the proton and anti-proton yields measured by
PHENIX~\cite{ppg026}. The neutral hadron contamination
is found to be negligible above $p_T = 5$~GeV/$c$ ($<1\%$).
The neutral photon-candidate spectra are corrected for conversions
removed by the charged contaminant subtraction with a $p_T$-independent
factor (5.9-7.3\% for different sectors based on simulation).  

\begingroup \squeezetable
\begin{table}[tbh]
\caption{\label{tab:syst}
Summary of the dominant sources of systematic errors on the
$\pi^0$ and inclusive $\gamma$ yields extracted independently with the
PbGl and PbSc electromagnetic calorimeters.  The error estimates are
quoted at two $p_{T}$ values in central events for the PbGl and PbSc.
For the combined $\pi^0$ and inclusive $\gamma$ spectra and
$\gamma/\pi^0$ ratios, the approximate statistical and systematical
errors are quoted for the most peripheral and most central reactions.}
\begin{ruledtabular} \begin{tabular}{lcccc}
 & \multicolumn{2}{c}{PbGl (Central)} &   \multicolumn{2}{c}{PbSc (Central)}\\
\multicolumn{1}{c}{$\pi^0$ error source} &\, 3\,GeV/$c$ & \,8.5\,GeV/$c$  &\,  3\,GeV/$c$ & \, 8.5\,GeV/$c$  \\
\hline
Yield extraction &   8.7\%  &  7.0\%     &  9.8\% &   7.2\% \\
Yield correction &  12.1\%   &  12.0\%    & 10.3\%  & 12.5\% \\
Energy scale     &  13.8\% &   14.1\% &    10.5\% &   11.4\% \\
\hline
Total systematic&  20.3\% &   19.8\% &   17.7\% &   18.4\% \\
\hline
Statistical     &  10.6\% &   32.5\%   &   2.1\% &   10.5\% \\
\hline
\hline
Inclusive $\gamma$ error       & & &  &\\
\hline
Non-$\gamma$ correction   &  2.4\% &   2.4\%  &   3.2\% &   3.2\% \\
Yield correction          & 10.2\% &   12.0\% &   9.1\% &   12.5\% \\
Energy scale              & 15.7\% &   13.7\% &   12.4\% &   10.8\% \\
\hline
Total systematic          & 18.9\% &   18.4\% &  15.7\% &   16.8\% \\
\hline
Statistical               & 1.2\% &   14.1\% &    0.6\%  &     4.1\% \\
\hline\hline
\quad $\gamma/\pi^0$ syst. & 13.6\% &  12.6\% &   14.0\% &   13.4\% \\
\quad $\gamma/\pi^0$ stat. & 10.7\% &   35.4\% &   2.2\%  &   11.3\% \\
\hline
\hline
\multicolumn{5}{c}{Total errors PbGl and PbSc combined} \\
\hline
& \multicolumn{2}{c}{Peripheral}   &  \multicolumn{2}{c}{Central} \\
\multicolumn{1}{c}{Error} &\,3\,GeV/$c$ & \,8.5~GeV/$c$  &\,  3\,GeV/$c$ & \,8.5\,GeV/$c$  \\
\hline
\quad $\pi^0$ syst. & 13.2\% &   17.0\% &  13.9\% &   16.1\% \\
\quad $\pi^0$ stat. & 3.0\% &    35.3\%  &   1.8\% &   9.6\% \\
\hline
\quad $\gamma$ syst.  & 11.4\% &   15.6\% & 11.5\% &   15.9\% \\
\quad $\gamma$ stat.  & 3.0\% &    28.8\% & 0.6\% &    3.8\% \\
\hline
\quad $\gamma/\pi^0$ syst. & 9.9\% &   13.1\% &  9.7\% &   11.2\% \\
\quad $\gamma/\pi^0$ stat. & 4.2\% &    45.6 \% &  1.9\% &    10.3\% \\
\hline
\quad $\gamma/\pi^0$ bkg calc.   &   \multicolumn{2}{c}{4\%}        &   \multicolumn{2}{c}{4\%}       \\
\end{tabular} \end{ruledtabular}
\end{table}
\endgroup

The raw spectra are normalized to one unit of rapidity
and full azimuth (the purely geometrical acceptance correction
is $\sim 1/0.35$) . The spectra are further
corrected for (i) the detector response (energy resolution, dead
areas), (ii) the reconstruction efficiency (PID cuts), and (iii)
occupancy effects (cluster overlaps). These corrections are
quantified by embedding simulated single $\gamma$'s, $\pi^0$'s, or
$\eta$'s from a full PHENIX GEANT simulation into real
events, and analyzing the merged events with the same analysis cuts
used to obtain the real yields. 
The overall $\pi^0$ yield correction
was $\sim$2.5 with a centrality dependence of $\lesssim$25\%. The
losses were dominated by fiducial and asymmetry cuts. The nominal
energy resolution was adjusted in the simulation
by smearing the energies with a constant term of $\sim5\%$ for PbSc and
$\sim7\%$ for PbGl to
reproduce the measured width of the $\pi^0$ peak observed at each $p_T$. 
The shape, position, and width of the $\pi^0$ peak
measured for all centralities were confirmed to be well
reproduced by the embedded data.

The energy calibration of the
EMCal was corroborated by the position of the $\pi^0$ invariant mass
peak, by the energy deposit from minimum ionizing charged particles
traversing the EMCal (PbSc), and by the correlation between the
measured momentum of electron and positron tracks identified by the
ring-imaging Cherenkov detector and the associated energy deposit in
the EMCal. From these studies it is determined that the accuracy of the energy
measurement was better than 1.5\%.

The main sources of systematic errors in the PbSc and PbGl
measurements are  the uncertainties in: (i) the yield
extraction, (ii) the yield correction, and (iii) the energy scale. The
relative contributions of these effects to the total error differ for
the PbSc and PbGl (Table~\ref{tab:syst}).  The weighted average of the
two independent measurements reduces the total error.
The final systematic errors on the spectra are at the level of $\sim 15-20\%$
(Table~\ref{tab:syst}). A
correction for the true mean value of the $p_T$ bin is applied to the steeply
falling spectra.

\begin{figure}[htb]
\includegraphics[width=1.0\linewidth]{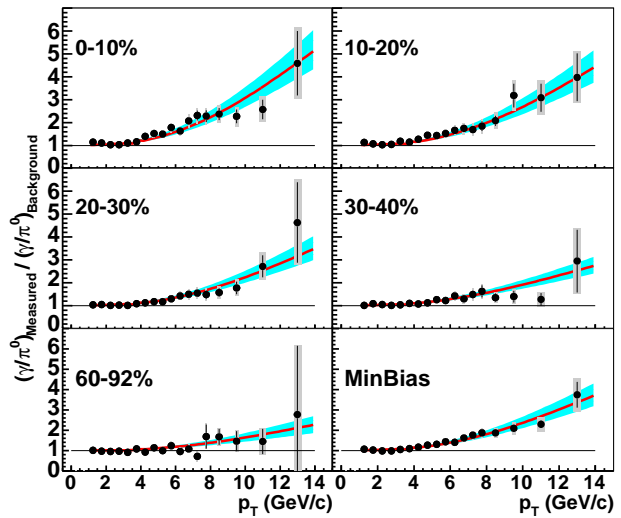}  
\caption{\label{fig:gamma_pi_ratios} 
Double ratio of measured
$(\gamma/\pi^0)_{\mathrm{Measured}}$ invariant yield ratio to the background decay
$(\gamma/\pi^0)_{\mathrm{Background}}$ ratio as a function of $p_T$ for minimum bias
and for five centralities of Au+Au
collisions at $\sqrt{s_{_{NN}}}$ = 200~GeV (0-10\% is the most
central). Statistical and total errors are indicated separately on each
data point by the vertical bar and shaded region, respectively.
The solid curves are the ratio of pQCD predictions described in the text
to the background photon
invariant yield based on the measured $\pi^0$ yield 
for each centrality class. The shaded region around the curves
indicate the variation of the pQCD calculation for scale changes from $p_T/2$
to $2p_T$, plus the $\langle N_{coll}\rangle$ uncertainty.}
\end{figure}

The completely corrected and combined PbSc and PbGl
inclusive photon yields are compared to the
expected yields of background photons
from hadronic decays in Fig.~\ref{fig:gamma_pi_ratios} for minimum bias
Au+Au collisions and for five centrality bins.
The decay photon calculations are based
on the measured $\pi^0$ and $\eta$ spectra~\cite{ppgtbp}
assuming $m_T$-scaling for
all other radiative decays ($\eta'$,$K^0_s$,$\omega$). The comparison
is made as the ratio of measured (inclusive) 
$\gamma/\pi^0$ and calculated background 
$\gamma/\pi^0$ since this has the
advantage that many uncertainties, such as the energy scale, cancel
to varying extent in the ratio.
Since the $\pi^0$ spectra of the background calculations
are taken to be the same as the measured spectra we have
\begin{equation}
R_\gamma = \frac{\left(\gamma/\pi^0\right)_{\mathrm{Measured}}}
{\left(\gamma/\pi^0\right)_{\mathrm{Background}}} \approx
\frac{\gamma_{\mathrm{Measured}}}{\gamma_{\mathrm{Background}}}
\label{eq:double_ratio}
\end{equation}
and any significant deviation of the double ratio above unity
indicates a direct photon excess.
In Fig.~\ref{fig:gamma_pi_ratios}
an excess is observed at high $p_T$ with a  magnitude that
increases with increasing centrality of the collision. 

\begin{figure}[htb]
\includegraphics[width=1.0\linewidth]{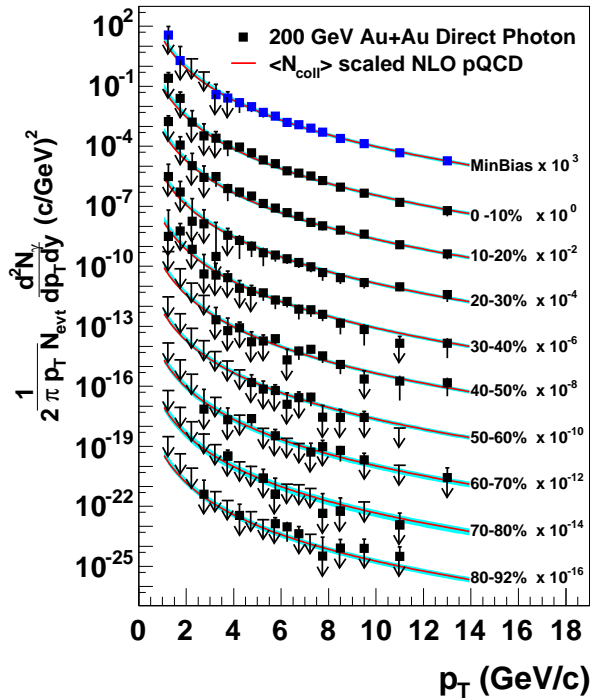}  
\caption{\label{fig:gamma_spectra}
Direct $\gamma$ invariant yields as a function of transverse momentum
for 9 centrality selections and minimum bias
Au+Au collisions at $\sqrt{s_{_{NN}}}$
= 200~GeV.  The vertical error bar on each point indicates the total error.
Arrows indicate measurements consistent with zero yield with the tail of
the arrow indicating the 90\% confidence level upper limit.
The solid curves are pQCD
predictions described in the text.
}
\end{figure}

The measured results are compared 
to NLO pQCD predictions~\cite{Gordon:1993qc}, scaled by the number of binary
nucleon collisions for each centrality selection. 
The same calculations are in agreement with the PHENIX
direct photon measurement~\cite{ppg049} for $p+p$ collisions
at the same $\sqrt{s}$,
and similar NLO pQCD calculations
provide a good description of the measured $\pi^0$
production in $p+p$ collisions~\cite{ppg024}.  
The calculations were performed~\cite{Gordon:1993qc,ppg049}
with normalization and factorization scales equal to $p_T$,
the CTEQ6~\cite{Pumplin:2002vw} set of parton distribution functions, and
the GRV set of fragmentation functions~\cite{Gluck:1992zx}.
The direct photon spectra extracted as
$\gamma_{\mathrm{Direct}}=(1-R^{-1}_{\gamma})\cdot \gamma_{\mathrm{Measured}}$
are shown in Fig.~\ref{fig:gamma_spectra} for all nine centrality
selections as well as minimum bias, and compared to the same NLO 
calculations. 
The binary collision scaled predictions
are seen to provide a good description of the measured direct
photon spectra (Fig.~\ref{fig:gamma_spectra}). 
The increasing ratio with centrality seen in Fig.~\ref{fig:gamma_pi_ratios}
is therefore attributed to the decreasing decay background due
to $\pi^0$ suppression~\cite{ppg014}.

Medium effects in $AA$
collisions are often presented using
the \emph{nuclear modification factor} given as the
ratio of the measured $AA$ invariant yields to the $NN$-collision-scaled 
$p+p$ invariant yields:
\begin{equation}
R_{AA}(p_T)\,=\,\frac{(1/N^{evt}_{AA})\,d^2N_{AA}/dp_T dy}{\langle N_{coll}\rangle/\sigma_{pp}^{inel} \,\times\,
d^2\sigma_{pp}/dp_T dy},
\label{eq:R_AA}
\end{equation}
where the $\langle N_{coll}\rangle/\sigma_{pp}^{inel}$ is the
average nuclear thickness function, $\langle T_{AA} \rangle$,
in the centrality bin under consideration (Ref~\cite{ppg014}).
$R_{AA}(p_T)$ measures the deviation of $AA$ data from an incoherent
superposition of $NN$ collisions.

\begin{figure}
\includegraphics[width=1.0\linewidth]{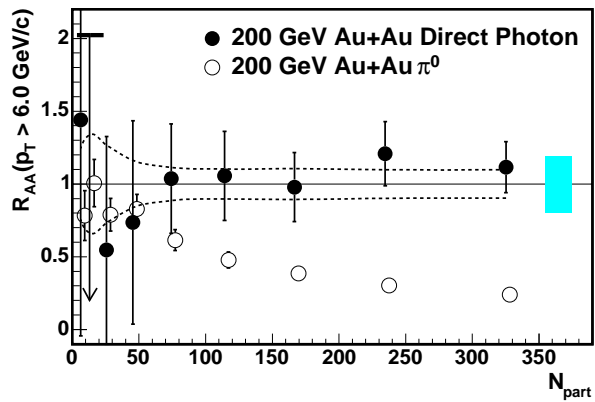}
\caption{Ratio of Au+Au yield to $p+p$ yield normalized by the number of
binary nucleon collisions as a function of centrality given by
$N_{part}$ for direct $\gamma$ (closed circles) and
$\pi^0$ (open circles) yields
integrated above 6~GeV$/c$. The $p+p$ direct photon yield is taken as the
NLO pQCD prediction described in the text.
The error bars indicate the total error excluding the error on 
$\langle N_{coll} \rangle$ shown by the dashed lines and the scale uncertainty of the NLO calculation shown by the shaded region at the right.
}
\label{fig:photon_R_AA}
\end{figure}


The centrality dependence of the high $p_T$ $\gamma$ production represented 
as a function of the number of participating nucleons,
$N_{part}$, is shown by the closed circles in
Fig.~\ref{fig:photon_R_AA}.  The production in Au+Au collisions
relative to $p+p$ is characterized by the $R_{AA}(p_T>6$ GeV$/c)$ ratio of
Eq.~(\ref{eq:R_AA}) as the ratio of Au+Au over the 
$\langle N_{coll}\rangle$-scaled $p+p$
yields each integrated above 6~GeV/$c$.  The direct photon $p+p$ yields
are taken as the NLO pQCD predictions described above.  As noted
above, the high $p_T$ direct $\gamma$ production is observed to scale
as the 
$\langle N_{coll}\rangle$-scaled $p+p$ $\gamma$ yield prediction for all
centralities. This is in sharp contrast~\cite{ppg014} to the centrality
dependence of the $\pi^0$ $R_{AA}(p_T>6$ GeV$/c)$ shown by open circles in
Fig.~\ref{fig:photon_R_AA} where
the measured $\pi^0$ yield~\cite{ppg024} is used as the $p+p$
reference in Eq.~(\ref{eq:R_AA}).

The observed close agreement between
the measured yields and NLO calculations is in striking contrast
to observations for central Pb+Pb collisions at $\sqrt{s_{_{NN}}}$ =
17.3~GeV~\cite{Aggarwal:2000th} where the measured photon 
yield exceeds the
$\langle N_{coll}\rangle$-scaled $p+p$ yield by about a factor of two.
The present result constrains modifications of the initial parton distributions, 
or of the fragmentation contributions~\cite{Fries:2002kt,Zakharov:2004bi} 
(in these NLO calculations the contribution 
is significant: $\sim 50\%$ at 3.5 GeV/$c$ and $\sim 35\%$ at
10 GeV/$c$), or additional photon yield from thermal radiation 
to levels comparable to the present measurement uncertainty.

%

In summary, the transverse momentum spectra of direct photons have
been measured at mid-rapidity up to $p_T\approx$ 13~GeV$/c$ for nine
centrality bins of Au+Au collisions at $\sqrt{s_{_{NN}}}$ =
200~GeV. The significance of the direct photon signal increases with
collision centrality due to the increasingly suppressed $\pi^0$
production and associated decrease in the photon background from
hadron decays.  The direct photon spectral shapes and invariant yields
are consistent with NLO pQCD predictions for $p+p$ reactions scaled by
the average number of inelastic $NN$ collisions for each centrality
class. The close agreement between measurement and the binary scaled
pQCD predictions of the direct photon yield suggests that nuclear
modifications of the quark and gluon distribution functions in the relevant
region of momentum fraction $x$ are minor. The result provides strong
confirmation that the observed large suppression of high $p_T$ hadron
production in central Au+Au collisions is dominantly a final-state
effect due to parton energy loss in the dense produced medium, rather
than an initial-state effect.

%


We thank the staff of the Collider-Accelerator and Physics
Departments at BNL for their vital contributions.  We acknowledge
support from the Department of Energy and NSF (U.S.A.), 
MEXT and JSPS (Japan), CNPq and FAPESP (Brazil), NSFC (China), 
CNRS-IN2P3 and CEA (France), 
BMBF, DAAD, and AvH (Germany), 
OTKA (Hungary), DAE and DST (India), ISF (Israel), 
KRF and CHEP (Korea), RMIST, RAS, and RMAE (Russia), 
VR and KAW (Sweden), U.S. CRDF for the FSU, 
US-Hungarian NSF-OTKA-MTA, and US-Israel BSF.



\end{document}